\documentclass[a4paper]{spie}  %>>> use this instead for A4 paper
\usepackage[]{graphicx}

\title{Specsim: The MIRI Medium Resolution Spectrometer Simulator} 

\author{Nuria P. F. Lorente*\supit{a},
        Alistair C. H. Glasse\supit{a},
        Gillian S. Wright\supit{a}  and 
        Macarena~Garc\'{i}a-Mar\'{i}n\supit{b}
\skiplinehalf
\supit{a}UK Astronomy Technology Centre, Royal Observatory,
  Blackford Hill, Edinburgh  EH9 3HJ, United Kingdom \\
\supit{b}Dpto. Astrofisica Molecular e Infrarroja,
Instituto de Estructura de la Materia, C/. Serrano 121,
28006 Madrid, Spain}

\authorinfo{* n.lorente@roe.ac.uk; previously N.\ P.\ F.\ M${\mathrm{ ^c}}$Kay; www.roe.ac.uk}
\begin{document} 
  \maketitle 

%%%%%%%%%%%%%%%%%%%%%%%%%%%%%%%%%%%%%%%%%%%%%%%%%%%%%%%%%%%%% 
\begin{abstract}
MIRI, the Mid-InfraRed Instrument, is one of four instruments being built 
for the James Webb Space Telescope, and is developed jointly between an European
Consortium and the US. 
In this paper we present a software data simulator for one of MIRI's four
instruments: the Integral Field Unit (IFU) Medium Resolution Spectrometer
(MIRI-MRS), the first mid-infrared IFU spectrograph, and one of the first IFUs
to be used in a space mission.
To give the MIRI community a preview of the properties of 
the MIRI-MRS data products before the telescope is operational, the 
Specsim tool has been developed to model, in software, the operation of 
the spectrometer. 
Specsim generates synthetic data frames approximating 
those which will be taken by the instrument in orbit. The program models 
astronomical sources and generates detector frames using the 
predicted and measured optical properties of the telescope and MIRI. These frames can 
then be used to illustrate and inform a range of operational activities, 
including data calibration strategies and the development and testing of 
the data reduction software for the MIRI-MRS. Specsim will serve as a 
means of communication between the many consortium members by providing a 
way to easily illustrate the performance of the spectrometer under 
different circumstances, tolerances of components and design scenarios.

\end{abstract}

%>>>> Include a list of keywords after the abstract 

\keywords{JWST/MIRI, simulations, spectroscopy: integral field, 
telescopes: jwst, data: modelling, instrument simulation}

%%%%%%%%%%%%%%%%%%%%%%%%%%%%%%%%%%%%%%%%%%%%%%%%%%%%%%%%%%%%%
\section{INTRODUCTION}
\label{sec:intro}  % \label{} allows reference to this section

\begin{figure}[t]
  \begin{minipage}[c]{0.49\linewidth}
  \begin{center}
    \resizebox{\linewidth}{!}{
      \includegraphics{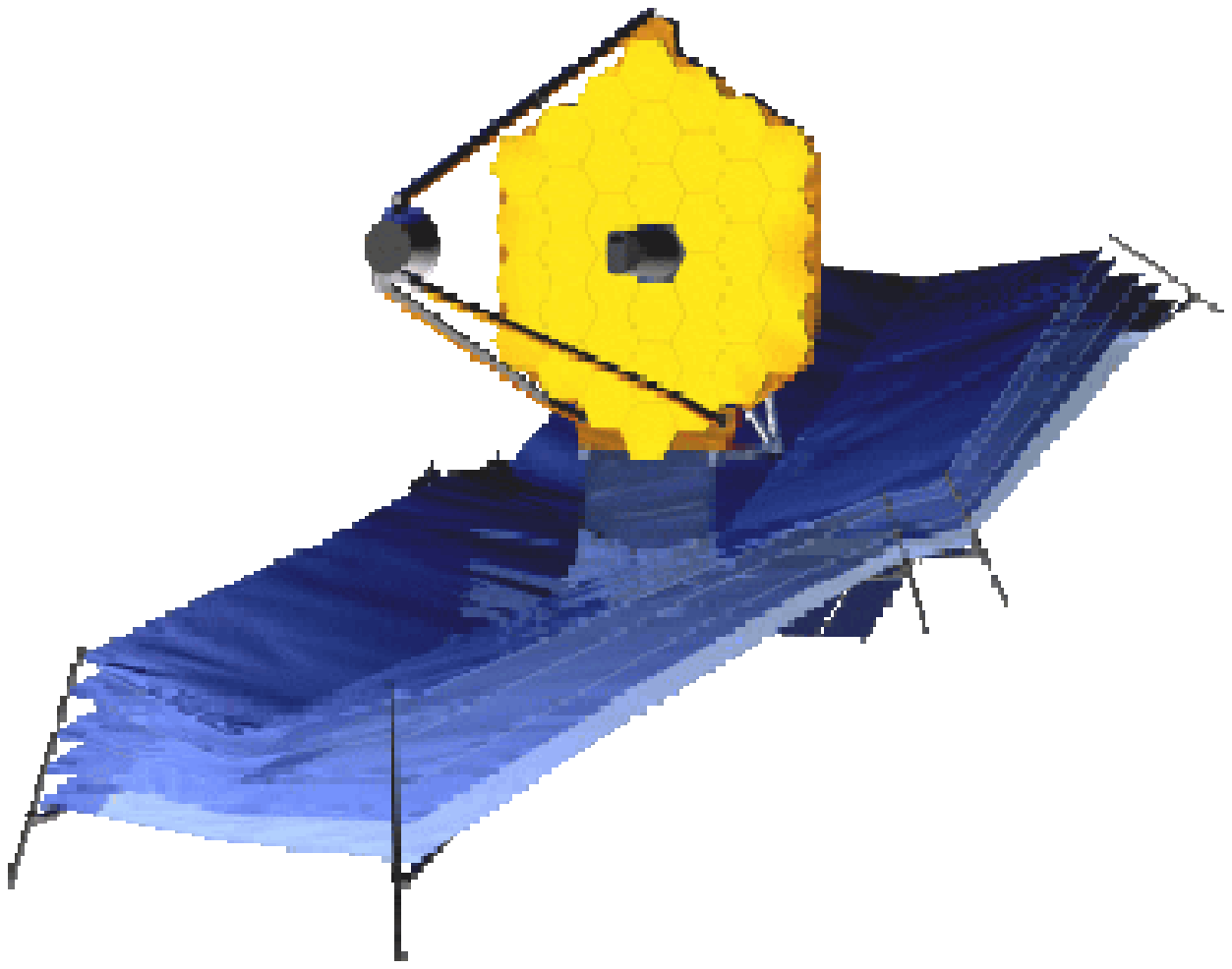}
    }
  \end{center}
  \end{minipage}
  \begin{minipage}[c]{0.49\linewidth}
  \begin{center}
    \resizebox{\linewidth}{!}{
      \includegraphics{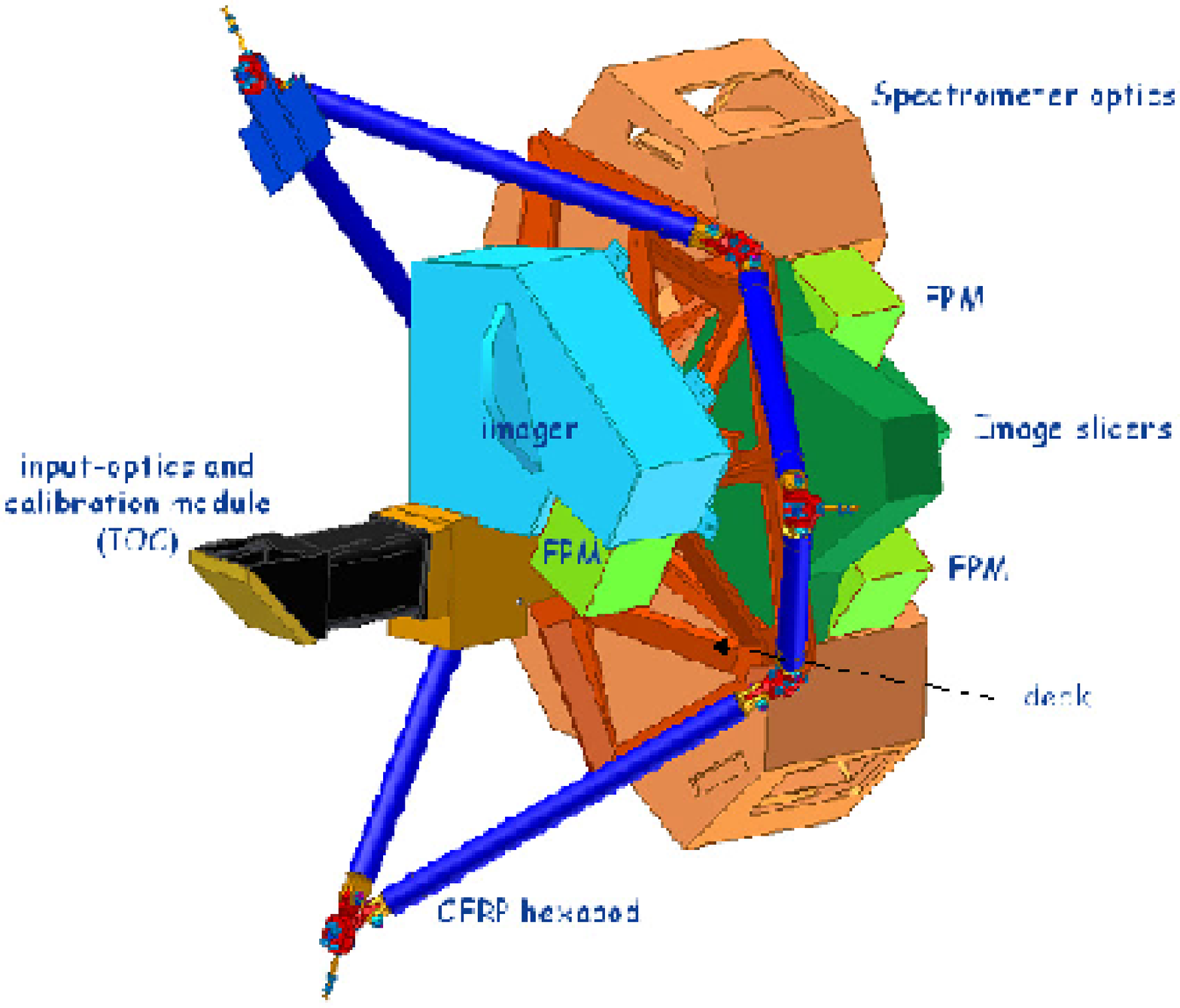}
    }
  \end{center}
  \end{minipage}
  \caption{The James Webb Space Telescope (left) and the MIRI instrument (right).}
  \label{fig:jwst}
\end{figure}

MIRI, the Mid-InfraRed Instrument, is one of four instruments being built for the James Webb
Space Telescope (Figure~\ref{fig:jwst}), scheduled to be launched in 2013 and
placed in a semi-stable orbit at the second Lagrangian point.
The design and development of MIRI is being carried out as a collaborative project
between an European Consortium of 21 institutes from 10 countries, under the
auspices of ESA, and the US.
MIRI itself consists of four instruments: an imager, a coronograph, a
low-resolution spectrograph, and an Integral Field Unit (IFU) Medium Resolution
Spectrometer (MIRI-MRS). The latter will be the first IFU spectrograph to
operate in the mid-infrared, and one of the first such instruments to be used in
a space mission.

\begin{figure}[b]
  \begin{center}
    \resizebox{0.49\linewidth}{!}{
       \includegraphics{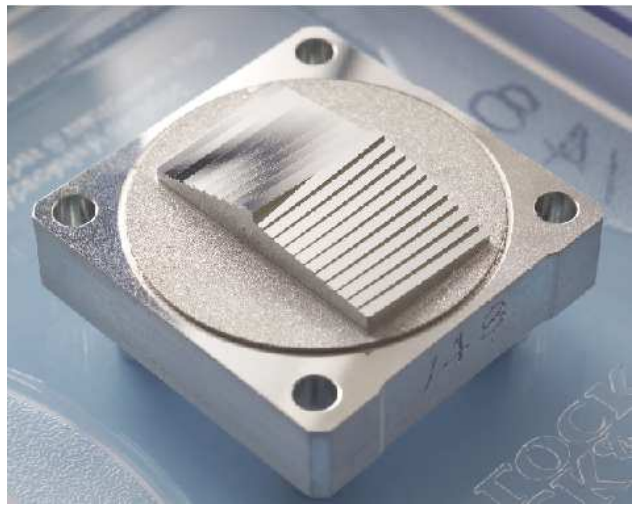}
    }
  \end{center}
  \caption{One of the four image slicers for the MIRI Medium Resolution
Spectrometer. The slicer shown is for the first spectral channel (4.87 -- 7.76
$\mu$m) which will divide the image plane into 21 slices and align these
sub-images with the spectrometer's input slit, in preparation for
dispersion. (Photograph credit: Cranfield University, UK).
}
  \label{fig:slicer}
\end{figure}

The MIRI-MRS is an integral field unit (IFU) spectrometer\cite{wrc+04}{}, which
allows spectroscopy to be carried out on a 2-dimensional area of sky in a single
observation. The primary component of the IFU is its image-slicing mirror\cite{wlo+06}{}
(Figure~\ref{fig:slicer}), which divides the rectangular field of view 
into a number of narrow slices. These are then arranged along the entrance slit
of a first-order diffraction grating, which carries out the dispersion.

\begin{figure}[t]
  \begin{center}
    \resizebox{0.8\linewidth}{!}{
       \includegraphics{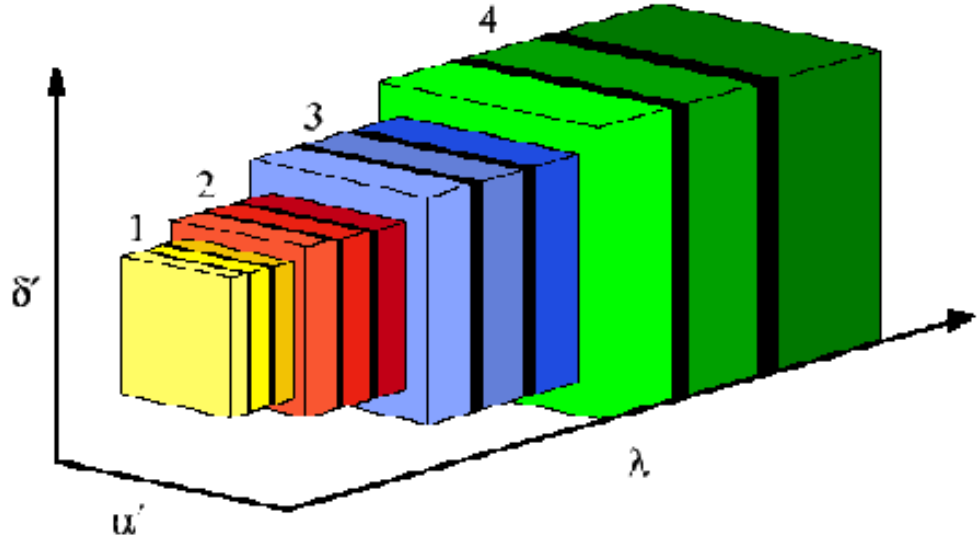}
    }
  \end{center}
  \caption{The field-of-view and bandwidth coverage (to scale) of MIRI-MRS channels 1 to 4
  The gradation of colour from light
  to dark in each channel denotes the three exposures A, B and C. The black
  bands between channels are those spectral regions 
  where the wavelength coverage of adjacent exposures overlap.}
  \label{fig:channels}
\end{figure}

The spectrometer operates in the spectral range 5--28~$\mu$m, with a resolution of
${\mathrm{R}\sim3000}$. The band is divided into 4 IFU channels, which
are observed simultaneously. Each channel is equipped with an IFU image slicer
designed to match the width of each slice to the diffraction-limited
point-spread function of the telescope, at the wavelength of each channel.
An observation over the entire spectral band is carried out in a set of three
exposures, the spectral band of each IFU channel being subdivided into 3
sub-bands by means of dichroic filters.  
A diagram showing the relative field of view and spectral bandwidth of each channel and exposure is
shown in Figure~\ref{fig:channels}.

For each of the three exposures, the data are
captured on one of two $1024\times1024$ pixel detectors (one for each pair of
image-slicer channels: (1A, 2A), (3A, 4A), (1B, 2B) {\em etc}.
The expected sensitivity of the instrument is $1.2×10^{-20}$~Wm$^{-2}$ at
6.4~$\mu$m, and $5.6×10^{-20}$~Wm$^{-2}$ at 22.5~$\mu$m, and its spatial field
of view widens with increasing channel number, as shown in
Figure~\ref{fig:channels}. The main functional parameters for the MIRI-MRS are
summarised in Table~\ref{tab:mrs_parameters}.

\begin{table}[b]
\label{tab:mrs_parameters}
\caption{Summary of the MIRI-MRS parameters}
\begin{center}
\begin{tabular}{|cccccc|}
\hline
Channel & FoV & Slices & $\lambda$ & $\mathrm{R_{spectral}}$ & Exposure\\
& (arcsec$^2$) & & ($\mu$m) & & \\
\hline
  &                  &     & 4.87 -- 5.82  &              & A \\
1 & 3.70$\times$3.70 & 21  & 5.62 -- 6.73  & 2450 -- 3710 & B \\
  &                  &     & 6.49 -- 7.76  &              & C \\
\hline
  &                  &     & 7.45 -- 8.90  &              & A \\
2 & 4.70$\times$4.51 & 17  & 8.61 -- 10.28 & 2480 -- 3690 & B \\
  &                  &     & 9.94 -- 11.87 &              & C \\
\hline
  &                  &     & 11.47 -- 13.67 &              & A \\
3 & 6.20$\times$6.13 & 16  & 13.25 -- 15.80 & 2510 -- 3730 & B \\
  &                  &     & 15.30 -- 18.24 &              & C \\
\hline
  &                  &     & 17.54 -- 21.10 &              & A \\
4 & 7.74$\times$7.93 & 12  & 20.44 -- 24.72 & 2070 -- 2490 & B \\
  &                  &     & 23.84 -- 28.82 &              & C \\
\hline
\end{tabular}
\end{center}
\end{table}

The science goals of the MIRI-MRS encompass a broad area of study\cite{g+05}{},
including the formation and evolution of galaxies, the life-cycle of stars and
stellar systems, the study of molecular clouds as the focus for star and planet
formation, and the investigation of planetary evolution conducive to life.

\begin{figure}[t]
\begin{center}
\resizebox{!}{0.85\textheight}{
\includegraphics{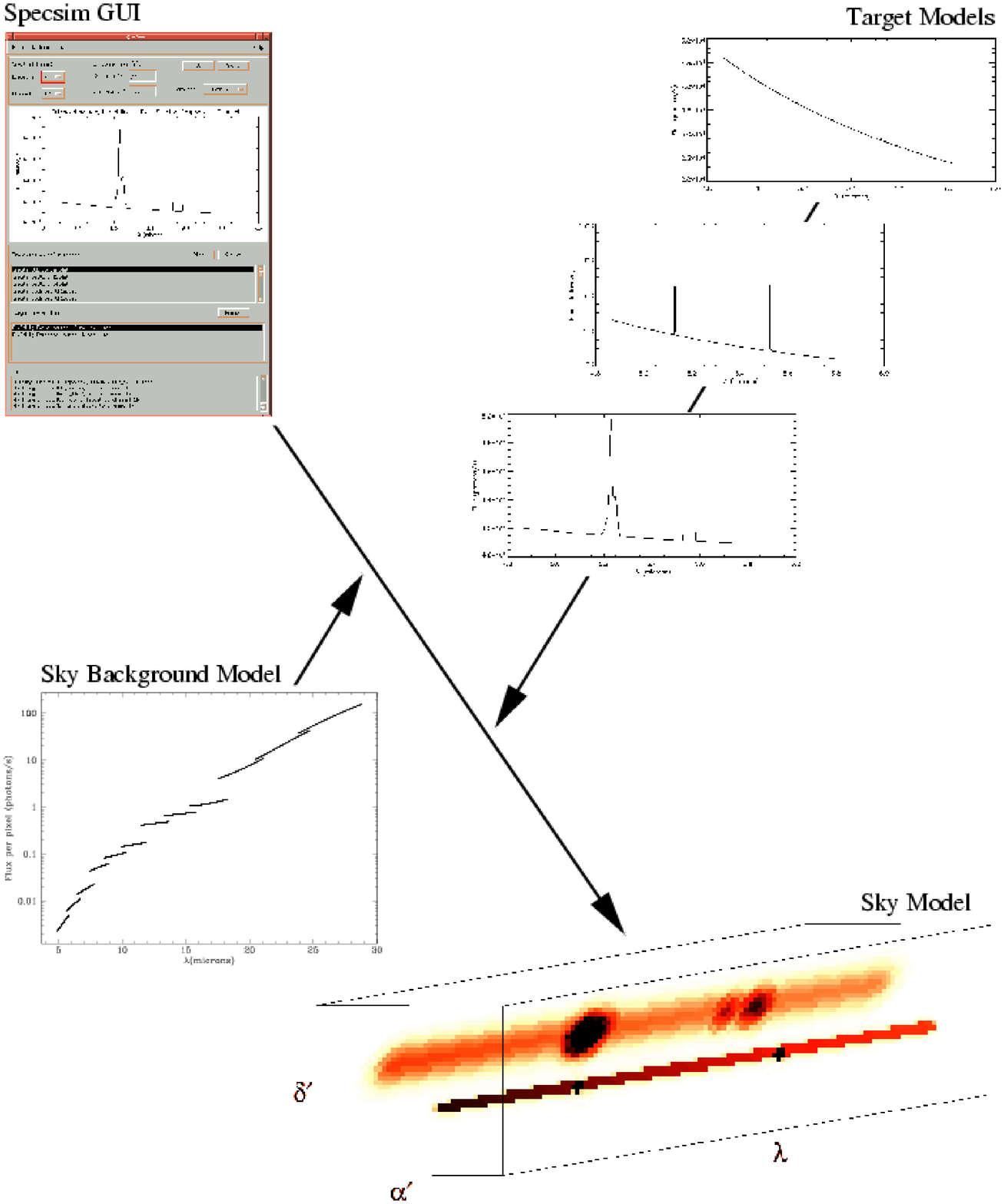}
}
\end{center}
\caption{Schematic showing the process of generating the Sky Model
(see Section~\ref{sec:skycube}). The user specifies the required observing
parameters (channel, exposure, exposure time, data quality) and loads previously
created astronomical target definition files using Specsim's user interface
(top-left). When the simulation is started, Specsim creates a model of the
field-of-view (the Sky Model, bottom-right) using its internal model of the sky
background (centre-left) and the spatial and spectral target definitions
imported by the user (top-right). As can be seen in the Sky Model (bottom-left)
this particular simulation models a point-source in continuum with two narrow
spectral lines, and an extended source with a broad line and two narrow spectral
lines. 
}
\label{fig:skycube}
\end{figure}

\clearpage
%%%%%%%%%%%%%%%%%%%%%%%%%%%%%%%%%%%%%%%%%%%%%%%%%%%%%%%%%%%%%
\section{SPECSIM: The MIRI-MRS Software Simulator}

Specsim, a software application developed in 
IDL\footnote[2]{IDL is a registered trademark of Research Systems Inc. for their
Interactive Data Language software.}, generates FITS frames which
approximate the images which will be produced by the MIRI-MRS.
It provides the MIRI community with the opportunity to study the properties of
MRS-like data products before the telescope is operational in orbit.
Activities such as the design and development of data reduction and analysis
software, which require an understanding of the MIRI-MRS data to inform the
design process as well as access to datasets with wich to test the software
tools subsequently developed, can greatly benefit from the availability of
simulated data.
Similarly, the development of calibration strategies relies on the understanding
of the properties of the instrument and their effect on the astronomical data
images, and therefore the quality of calibration strategies can be improved if
good data models are available.

The process of modelling MIRI-MRS data is carried out by Specsim in two main
stages: First, a model of the instrument's field-of-view is created in three
dimensions (two spatial and one spectral). 
This model is then used to generate a detector image, using Specsim's internal
model of the JWST and the MIRI-MRS.

\subsection{Modelling the Sky}\label{sec:skycube}
Specsim builds a model of the instrument's field-of-view, for each of the
spectrometer's four spectral channels and three exposures. The primary component
of this Sky Model is a number of user-defined astronomical target models,
generated by means of a set primitives, provided by the program, with which the
user may specify the morphological and spectral characteristics of each target
in the field.
Currently available target primitives include the following:

\subsubsection*{Target Morphology:}
\begin{itemize}
  \item Point Source: A point-source, characterised by the point-spread function of
        the instrument, centred on a user-defined position in (or outside of)
the field-of-view.
  \item Biaxial Gaussian: A Gaussian, with user-specified semi-major and
semi-minor FWHM and inclination, and centred on a user-specified position.
  \end{itemize}

\subsubsection*{Spectral Properties:}
\begin{itemize}
\item Continuum Properties:
   \begin{itemize}
   \item Black-body Continuum: A simple black-body continuum model, given by

$ F(\lambda) = F_{10}\left(\frac{\lambda_{10}}{\lambda}\right)^5 
   \frac{e^{\frac{hc}{\lambda_{10} k T}}-1}{e^{\frac{hc}{\lambda k T}}-1} $,

with a user-specified $10\mu$m flux ($F_{10}$) and temperature ($T$).

   \item Black-body Continuum + Dust: A good approximation of the behaviour of
   a hot dust component is achieved using a black-body function, modified by
   $\lambda^{-2}$:

$F(\lambda) = F_{10}\left(\frac{\lambda_{10}}{\lambda}\right)^5 
   \frac{e^{\frac{hc}{\lambda_{10} k T}}-1}{e^{\frac{hc}{\lambda k T}}-1} 
   \times \frac{1}{\lambda^2} $,

with the $10\mu$m flux ($F_{10}$) and temperature ($T$) specified by the user.

   \item Polynomial: A continuum model described by the polynomial function

$F(\lambda) = A + Bx + Cx^2 + Dx^3 $, 

where A, B, C and D are given by the user.

   \item SB99 Model (Young populations):
This primitive makes use of the Starburst99 models\cite{lsg+99, vl05}{}. 
The model is well sampled to $\sim$10 $\mu$m (restframe), and is recommended
for modelling galaxies at high redshifts. 
Primitives for starburst models for redshifts $z=0-10$ are available; 
by default Specsim uses a model corresponding to a young (6~Myr)
instantaneous burst of star formation, normalised to a total mass of
10$^6$M$_\odot$, with the Salpeter Initial Mass Function (IMF) and Geneva tracks
adapted for a galaxy at a redshift $z=6$. 

   \item SB99 Model (Intermediate populations):
As for the previous continuum type, this primitive makes use of the Starburst99
models\cite{lsg+99, vl05}{}. 
Starburst models at integer redshifts in the range $z=0-10$ may be selected, 
with the default model corresponding to a 50~Myr old instantaneous burst of star
formation, normalised to a total mass of 10$^6$M$_\odot$, with Salpeter IMF
and Geneva tracks adapted for a galaxy at redshift $z=6$.

   \item Bruzual \& Charlot (Old populations):
This makes use of the models by Bruzual \& Charlot\cite{bc03}{}. The model is 
well sampled to $10 \mu{}m$ (restframe).
The default model corresponds to 2~Gyr old instantaneous burst of star
formation, normalised to a total mass of $10^6$M$_\odot$
with Salpeter IMF and Padova tracks
adapted for a galaxy at redshift $z=6$. As above, starburst models at redshifts
in the range $z=0-10$ may also be selected. 

   \item Continuum \& PaH Model:
This model, developed by Lagache et al.\cite{ldp+04}{}, is valid in the wavelength range
$\lambda = 4\mu{}$m -- 3 mm.
The default model used by Specsim corresponds to an ULIRG of
luminosity $L = 5\times 10^{12} L_\odot$, located at redshift $z=6$, but models
in for redshifts $z=1-10$ are also available.
   \end{itemize}

\item Broad Spectral Features:
  \begin{itemize}
  \item Gaussian: A spectral feature defined by the Gaussian function, with a
user-specified central wavelength, peak flux and FWHM.

  \item $9.7 \mu$m Silicate: This primitive will load a normalised template of
  the $9.7\mu{}$m silicate feature and scale it to the requested peak flux.

  \item $18 \mu$m Silicate: This will load a normalised template of the silicate
  feature at $18\mu{}$m and scale it to the required peak flux.

  \end{itemize}
\item Narrow Spectral Lines: Unresolved spectral lines, in emission or
absorption, may be defined by specifying the required central wavelength of the
line and the line strength. Specsim will then use its internal instrument data
tables to calculate the appropriate line width, depending on the requested wavelength.
\end{itemize}

The above primitives may be used individually or in combination, to construct target
fields of arbitrary complexity. The target definitions are imported into
Specsim at run-time, by means of a previously created text file, and
%incorporated into the sky model.
%The user-specified target model, 
together with Specsim's internal models of the
zodiacal background, contributions from the telescope's solar shield, {\em
etc.}, are used to generate a target simulation cube, representing the
instrument's field of view over each channel's spectral range. This process is
illustrated in Figure~\ref{fig:skycube}.

\subsection{Modelling the MIRI-MRS}
\begin{figure}[t]
\begin{center}
\resizebox{!}{0.85\textheight}{
\includegraphics{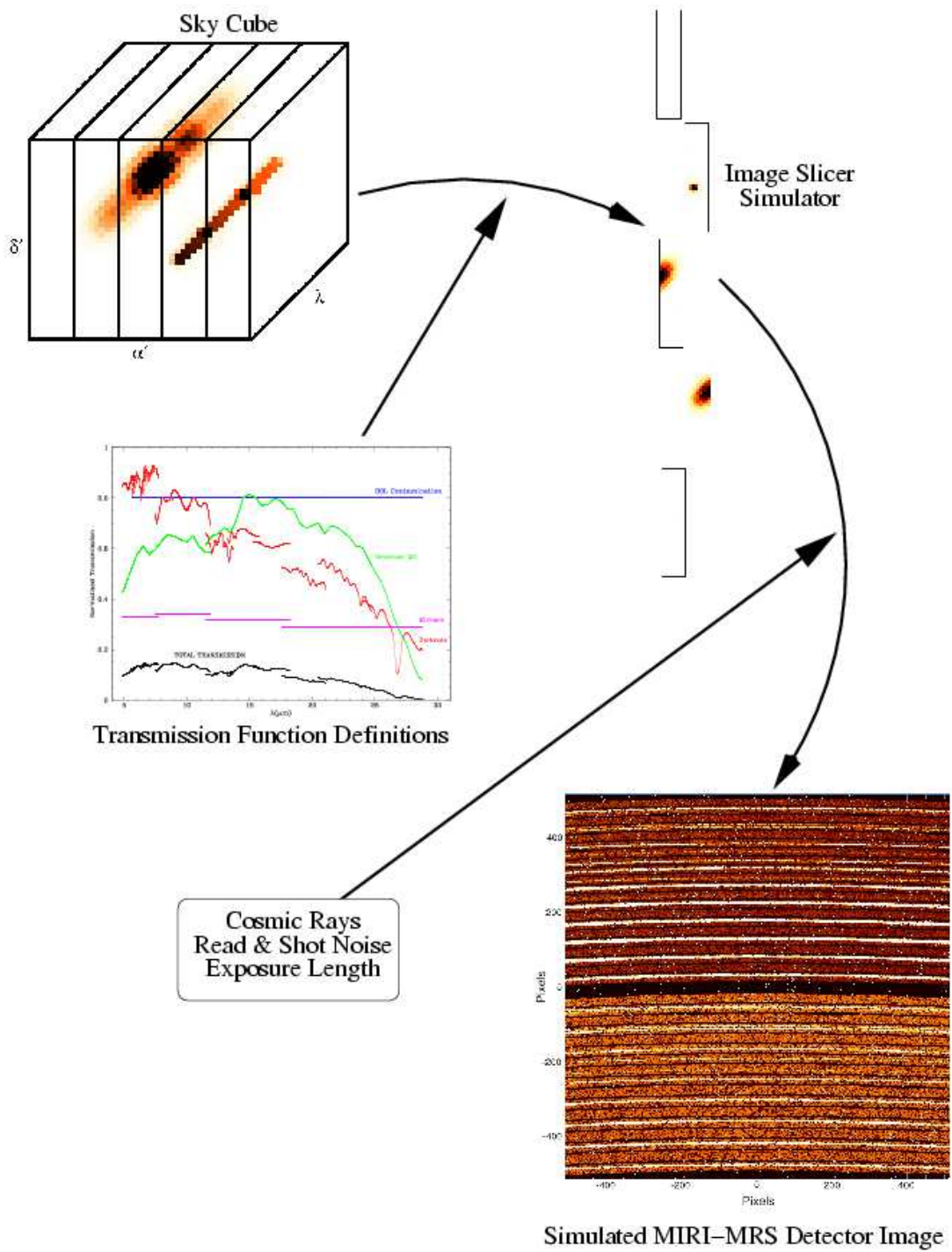}
}
\end{center}
\caption{Schematic of the process of generating a simulated detector frame from
the Sky Model. The standard or user-provided transmission functions
describing the instrument components are loaded and applied to the Sky Model data
(centre-left and top-left). The Sky Model is sliced in the spatial dimensions, with
the number of slices depending on the spectrometer channel being modelled, and
the slices are aligned as for input into the diffraction grating
(top-right). Dispersion by the spectrometer is then carried out on the vertically
aligned slices, to produce a detector image. Finally, the effects of cosmic rays,
integration and noise are implemented on the image, and
the simulated spectra for each pair of channels are
mapped onto each of two $1024\times1024$ pixel detectors (bottom-right).}
\label{fig:detimg}
\end{figure}

Once a model of the target field of view has been produced, Specsim simulates
the function of the spectrograph, producing a simulated spectroscopic
observation of the field. Figure~\ref{fig:detimg} shows a schematic of this process.

The MIRI-MRS model is generated in several stages. First, the contribution to
the observed flux by the instrument's optics and electronics is applied to the
sky model.
The behaviour of the instrument's components is described by means of
transmission functions, consisting of ASCII files listing the normalised
transmission of the subsystem (the efficiency of the mirrors, the behaviour of
the detectors, the performance of the dichroics, {\em etc.}) at each wavelength
across the band. The values for the transmission functions may be derived from
models of the behaviour of the components or be actual measurements taken during
laboratory testing of the individual subsystems.
The standard transmission functions are normally loaded automatically by the
program. In Specsim's expert mode, however, the transmission functions may be
provided by the user, affording extra flexibility in modelling specific aspects
of the instrument, simulating faults, and allowing the effect of specific
instrument characteristics on the final spectrometer image to be closely
studied.

The next step in the process is to simulate the effect of the image slicer,
which is done by assigning each pixel in the sky model to an IFU slice,
depending on its location in the field of view. Geometric deformations are then
applied, simulating the optical path of the telescope, the spectrometer's
pre-optics and the image-slicing mirror.

Dispersion of the image slices is then carried out, by mapping each pixel in the
Sky Model onto the corresponding pixel in the detector image, depending on
wavelength. 

\clearpage

Once the detector image is constructed, on-sky integration is implemented
(length of exposure may be supplied by the user) and
cosmic rays are added. Cosmic rays have the effect of saturating one or more
pixels on the detector, and rendering those pixels invalid for the remainder of
the exposure. This results in detector pixels with effective integration
times between zero (a cosmic ray hit in the first sub-integration of the
exposure) and a full integration (no cosmic ray). 

Finally, photon noise and read-noise are added to the detector image, and both
the Sky Model and the detector image are provided as FITS files for the
user. This allows the detector frames to be processed, analysed and compared
with the input targets, thus providing an useful test for the development of the
MIRI data reduction software, testing calibration strategies and
observation planning.

%\clearpage

%%%%%%%%%%%%%%%%%%%%%%%%%%%%%%%%%%%%%%%%%%%%%%%%%%%%%%%%%%%%%
\acknowledgments     %>>>> equivalent to \section*{ACKNOWLEDGMENTS}       

MIRI draws on the expertise of the following organisations: Ames 
Research Center, USA; Astron, Netherlands Foundation for Research in Astronomy;
CEA Service d'Astrophysique, Saclay, France; Centre Spatial de Li\'{e}ge,
Belgium; Consejo Superior de Investigaciones Cient\'{i}ficas, Spain; Danish Space
Research Institute; Dublin Institute for Advanced Studies, Ireland; EADS
Astrium, Ltd., European Space Agency, Netherlands; UK; Institute d'Astrophysique
Spatiale, France; Instituto Nacional de T\'{e}cnica Aerospacial, Spain;
Institute of Astronomy, Zurich, Switzerland; Jet Propulsion Laboratory, USA;
Laboratoire d'Astrophysique de Marseille (LAM), France; Lockheed Advanced
Technology Center, USA; Max-Planck-Insitut f\"{u}r Astronomie (MPIA),
Heidelberg, Germany; Observatoire de Paris, France; Observatory of Geneva,
Switzerland; Paul Scherrer Institut, Switzerland; Physikalishes Institut, Bern,
Switzerland; Raytheon Vision Systems, USA; Rutherford Appleton Laboratory (RAL),
UK; Space Telescope Science Institute, USA; Toegepast-Natuurwetenschappelijk
Ondeszoek (TNO-TPD), Netherlands; UK Astronomy Technology Centre (UK-ATC);
University College, London, UK; Univ. of Amsterdam, Netherlands; Univ. of
Arizona, USA; Univ. of Cardiff, UK; Univ. of Cologne, Germany; Univ. of
Groningen, Netherlands; Univ. of Leicester, UK; Univ. of Leiden, Netherlands;
Univ. of Leuven, Belgium; Univ. of Stockholm, Sweden, Utah State Univ. USA.

%%%%%%%%%%%%%%%%%%%%%%%%%%%%%%%%%%%%%%%%%%%%%%%%%%%%%%%%%%%%%
%%%%% References %%%%%

%\bibliography{journals,refs_miri}   %>>>> bibliography data in report.bib
%\bibliographystyle{spiebib}   %>>>> makes bibtex use spiebib.bst

\end{document}